\documentclass[letter]{ptptex}

\inst{$^{1}$Department of Physics, Shanghai University, Shanghai,
200444, P. R. China\\$^{2}$The Shanghai Key Lab of Astrophysics,
Shanghai, P. R. China\\$^{3}$Center of Theoretical Nuclear Physics,
National Laboratory of Heavy Ion Accelerator, Lanzhou 730000, P. R.
China\\$^{4}$Institute of Modern Physics, Lanzhou, 730000, P. R.
China}
\abst{We show that the field equation of Brans-Dicke gravity and
scalar-tensor gravity can be derived as the equation of state of
Rindler spacetime, where the local thermodynamic equilibrium is
maintained. Our derivation implies that the effective energy can not
feel the heat flow across the Rindler horizon.}

\begin{document}

\title{The equation of state for scalar-tensor gravity}
\author{Shao-Feng Wu$^{1,2}$\thanks{%
Corresponding author. Email: sfwu@shu.edu.cn; Phone: +86-021-66136202.},
Guo-Hong Yang$^{1,2}$, and Peng-Ming Zhang$^{3,4}$}
\maketitle

Motivated by black hole mechanics and Hawking radiation, it has become
increasingly clear that there is a deep connection between gravity and
thermodynamics. A key evidence of this connection was discovered by Jacobson
\cite{Jacobson}, who derived the Einstein equation as an equation of state
of local Rindler spacetime by constructing the equilibrium thermodynamics $%
\delta Q=TdS$. This derivation suggests a gravitational theory built upon
the principle of equivalence must be thought of as the macroscopic limit of
some underlying microscopic theory.

The relation between gravity and thermodynamics has been found also in other
spacetimes, including a general static spherically symmetric spacetime \cite%
{Padmanabhan}, the quasi-de Sitter geometry of inflationary universe \cite%
{Frolov}, the expanding universe \cite{Danielsson}, the quintessence
dominated accelerating universe \cite{Bousso}, and the spacetime with any
spatial curvature \cite{Cai}. The relation has been further disclosed in
gravity theories beyond Einstein gravity, including Lovelock gravity \cite%
{Cai,Cao}, braneworld gravity \cite{Cao1}, nonlinear gravity \cite%
{Eling,Akbar1,Cao}, and scalar-tensor gravity \cite{Akbar1,Cao,Akbar} etc.
However, in the nonlinear gravity and scalar-tensor gravity, it was argued
that the non-equilibrium thermodynamics instead of the equilibrium
thermodynamics should be taken into account to build the relation to gravity
\cite{Eling,Akbar,Cao}.

An alternative treatment to reinterpretate of the nonequilibrium correction
was introduced in \cite{Gong}, where a mass-like function is introduced to
connect the first law of thermodynamics and Friedmann equations in some
gravity theories. Recently, using the idea of \textquotedblleft
local-boost-invariance\textquotedblright\ introduced in \cite{Iyer}, as well
as a more general definition of local entropy, the nonlinear $f(R)$ gravity
field equation is derived from the Rindler space-time thermodynamics,
maintaining the local thermodynamic equilibrium \cite{Elizalde}. In the
present letter, we want to know whether the method developed in \cite%
{Elizalde} holds or not in more general scalar-tensor gravity. Compared with
the $f(R)$ gravity, there is the similar extended local entropy, but a new
degree of freedom, that is an additional scalar field mediating the
gravitational interaction.

Let us begin with briefly introducing (see the detail review in \cite%
{Jacobson1}) the local Rindler spacetime where we will construct the
thermodynamics. In the vicinity of any space-time point $p$, a free-falling
observer can use the equivalence principle to describe his local coordinate
system as flat. Then the observer can locally define a causal horizon as
follows. Choose a spacelike surface patch $B$ including $p$ and choose one
side of the boundary of the past of $B$. Near the point $p$, this boundary
is a congruence of the null geodesics orthogonal to $B$. These comprise the
horizon. The called "local Rindler horizon", is adjusted so that the
generators of the causal horizon have vanishing expansion $\theta $ and
shear $\sigma $.

Jacobson \cite{Jacobson} pointed out that we can construct equilibrium
thermodynamics of local Rindler horizon at $p$. The thermodynamics is
described by Clausius relation $\delta Q=TdS$. Here the entropy $S$ is
associated with the causal horizon, suggested by the observation that they
hide information. Jacobson suspected the entropy measuring the
\textquotedblleft many degrees of freedom outside\textquotedblright , what
presumably results in entanglement entropy just at the horizon. He then
assumed that the entropy is proportional to horizon area, namely the
proportionality between entropy and the horizon area, formulized by $S=\eta
A $. Jacobson takes $\eta $ as an unknown proportionality constant in
Einstein gravity. In $f(R)$ gravity, it is interesting to find that \cite%
{Elizalde}, through replacing the proportionality constant with a
field-dependent effective constant and using the boost-invariant truncation,
similar equilibrium thermodynamics can be constructed. The field-dependent
effective constant is motivated by a result that, for a static black hole in
$f(R)$ gravity, entropy can be re-expressed as the 1/4 area of the
bifurcation surface in units of an effective Newton constant \cite{Brustein}%
. In the present letter, we will also use field-dependent $\eta $ because
the black hole entropy in scalar-tensor gravity is also 1/4 of horizon area
in units of the corresponding effective Newton constant \cite{Cai2}.

The heat $Q$ is interpreted as the mean flux of (boost) energy across the
horizon. In Einstein gravity, Jacobson assumed that all the heat flow across
the horizon is the energy carried by matter. In $f(R)$ gravity, similar
perspective is adopted \cite{Elizalde}. For scalar-tensor gravity, here we
simply assume the heat flow across the horizon is the energy carried by
matter and scalar field. However, in the later of this letter, we will point
out that this definition of the energy (and heat) is a little coarse and an
important implication of this definition. To give the expression of heat
flux, we consider an approximate boost killing vector $\chi $ which is
future pointing on the causal horizon and vanishes at the space-time point $%
p $. Killing vector $\chi $ has a relation with the tangent vector of causal
horizon $K$: $\chi $ $=-k\lambda K+O(\lambda ^{2})$, where $k$ is the
acceleration of $\chi $ and $\lambda $ is the affine parameter of the
corresponding null geodesic line such that $\lambda =0$ at $p$. Thus, the
heat flux is given by $\delta Q=\int T_{ab}\chi ^{a}d\Sigma ^{b}$, where $%
T_{ab}$ denotes the energy--momentum tensor of matter and scalar field, and
the integral is taken over a small region of pencil of generators of the
horizon terminating at $p$. If the area element of horizon is $dA$ then $%
d\Sigma ^{b}=-K^{b}d\lambda dA$. Thus, the final expression for the
variation of heat, at leading order is%
\begin{equation}
\delta Q=-\int k\lambda T_{ab}K^{a}K^{b}d\lambda dA.  \label{dQ}
\end{equation}%
The quantum vacuum in flat spacetime for the generator of Lorentz boosts
could be treated as a Gibbs ensemble with temperature $T_{0}=1/2\pi $ in
units of Planck constant. Note that a uniformly accelerated observer behaves
as if immersed in a thermal bath at the Unruh temperature \cite{Unruh}%
\begin{equation}
T=kT_{0}=\frac{k}{2\pi }.  \label{T}
\end{equation}%
Now we consider the variation of entropy expression $S=\eta A$. Change in
the horizon area is given in terms of the expansion of the congruence of
null geodesics generating the horizon $\delta A=\int \theta d\lambda dA$.
Since the equation of geodesic deviation for null geodesic congruence is
given by the Raychaudhuri equation $d\theta /d\lambda =-\theta ^{2}/2-\sigma
_{ab}\sigma ^{ab}-R_{ab}K^{a}K^{b}$. Considering vanishing shear $\sigma $
and expansion $\theta $ terms in local Rindler coordinate system we get the
solution $\theta =-\lambda R_{ab}K^{a}K^{b}$ at leading order in $\lambda $.
Therefore, the relevant expression for the variation of entropy is evaluated
\begin{equation}
\delta S=-\lambda \int \left( \eta R_{ab}-\nabla _{a}\nabla _{b}\eta \right)
K^{a}K^{b}d\lambda dA.  \label{dS}
\end{equation}%
As noticed in \cite{Elizalde}, in this expression $\left( \eta ,K^{a}\nabla
_{a}\eta \right) $ are to be evaluated at its leading contribution in $%
\lambda $. Its boost-invariant part at first order in $\lambda $ has been
used to effectively incorporate the boost invariant notion of creating an
\textquotedblleft approximated bifurcation point at the first order in $%
\lambda $" at $p$ \cite{Iyer}. Considering the condition that Clausius
relation $\delta Q=T\delta S$ with heat (\ref{dQ}), temperature (\ref{T}),
and entropy (\ref{dS}) is satisfied for all vectors $K$, one can obtain%
\begin{equation}
T^{ab}=\frac{\eta }{2\pi }R^{ab}-\nabla ^{a}\nabla ^{b}\frac{\eta }{2\pi }%
+g^{ab}H,  \label{TR}
\end{equation}%
where the new term about $H$ is added since the tangent vector $K$
generating horizon has null norm. Hence, at this point we have a local
equation with two unknown functions $\left( \eta ,H\right) $. We will show
the functions may be determined by imposing the divergence free for the
energy--momentum tensor of matter.

Before going any further, it is time to introduce the concrete gravity
theory. We first consider the simple Brans-Dicke gravity. Its Lagrangian is
\cite{Brans}%
\begin{equation}
L=\phi R-\frac{w}{\phi }g^{ab}\partial _{a}\phi \partial _{b}\phi +L_{m},
\label{Lag}
\end{equation}%
where Brans-Dicke scalar field $\phi $ plays the role of the effective
gravitational constant ($\phi =\frac{1}{8\pi G_{eff}}$), and $L_{m}$ is the
Lagrangian of ordinary matter. Varying the action, we have the equation of
motion of scalar field%
\begin{equation}
2w\nabla ^{2}\phi -\frac{w}{\phi }\left( \nabla \phi \right) ^{2}+\phi R=0,
\label{em}
\end{equation}%
and gravitational field equation%
\begin{equation}
G_{ab}=\frac{1}{\phi }\left( T_{ab}^{m}+T_{ab}^{\phi }+T_{ab}^{e}\right) ,
\label{fe}
\end{equation}%
where%
\begin{equation}
T_{ab}^{\phi }=\frac{w}{\phi }\left[ \nabla _{a}\phi \nabla _{b}\phi -\frac{1%
}{2}g_{ab}\left( \nabla \phi \right) ^{2}\right] ,  \label{Tf}
\end{equation}%
\begin{equation}
T_{ab}^{e}=-g_{ab}\nabla ^{2}\phi +\nabla _{a}\nabla _{b}\phi .  \label{Te}
\end{equation}%
As aforesaid, we have simply assumed the heat flow across the horizon is the
energy carried by matter and scalar field. Now we will discuss it carefully.
It is clear that $T_{ab}^{m}$ denotes the energy--momentum tensor of matter
as it comes from the variation of the matter action with respect to metric.
But what is the energy-momentum tensor of scalar field? We only know when
the first term in (\ref{Lag}) vanishes, $T_{ab}^{\phi }$ is the
energy--momentum tensor of scalar field, since it comes from varying the
purely scalar field parts of the action with respect to metric, but do not
know how to explain the remained $T_{ab}^{e}$. In Ref. \cite{Santiago}, it
was argured that the appropriate definition of the scalar field
energy-momentum tensor should be formulated in Einstein frame. Obviously, it
is not helpful in our situation. One is tempted to identify all scalar field
terms $T_{ab}^{\phi }+T_{ab}^{e}$ with the energy-momentum tensor of the
scalar field by comparing the field equation (\ref{fe}) with Einstein field
equation. Naively, $T_{ab}^{e}$ can be explained as a kind of effective
energy. But we will show the unknown functions $\left( \eta ,H\right) $ and
the correct field equation (\ref{fe}) can be obtained if we take $%
T_{ab}^{\phi }$ as the energy--momentum tensor of scalar field instead of
the $T_{ab}^{\phi }+T_{ab}^{e}$.

Taking $T_{ab}^{\phi }$ as the energy--momentum tensor of scalar field, the
local equation (\ref{TR}) reads%
\[
T^{mab}+T^{\phi ab}=\frac{\eta }{2\pi }R^{ab}-\nabla ^{a}\nabla ^{b}\frac{%
\eta }{2\pi }+g^{ab}H.
\]%
Imposing the matter conservation $\nabla _{a}T^{mab}=0$, the above equation
can be expanded as%
\begin{equation}
\nabla _{a}T^{\phi ab}=\frac{\eta }{4\pi }\nabla ^{b}R+(\nabla _{a}\nabla
^{b}\nabla ^{a}-\nabla ^{b}\nabla ^{2})\frac{\eta }{2\pi }+\nabla ^{b}H-%
\frac{1}{2}(\nabla ^{2}\nabla ^{b}+\nabla _{a}\nabla ^{b}\nabla ^{a})\frac{%
\eta }{2\pi }.  \label{s1}
\end{equation}%
Setting $H=h+\nabla ^{2}\frac{\eta }{2\pi }$, Eq. (\ref{s1}) can be written
as%
\begin{equation}
\nabla _{a}T^{\phi ab}=\frac{\eta }{4\pi }\nabla ^{b}R+\nabla ^{b}h.
\label{s2}
\end{equation}%
Taking covariant divergence of $T_{ab}^{\phi }$ (\ref{Tf})%
\begin{eqnarray*}
\nabla _{a}T^{\phi ab} &=&\nabla _{a}\frac{w}{\phi }\left[ \nabla ^{a}\phi
\nabla ^{b}\phi -\frac{1}{2}g^{ab}\nabla ^{2}\phi \right] +\frac{w}{\phi }%
\left[ \nabla _{a}\nabla ^{b}\phi \nabla ^{a}\phi +\nabla ^{b}\phi \nabla
^{2}\phi -\nabla ^{b}\nabla ^{a}\phi \nabla _{a}\phi \right]  \\
&=&\frac{1}{2\phi }\nabla ^{b}\phi \left[ 2w\nabla ^{2}\phi -\frac{w}{\phi }%
\left( \nabla \phi \right) ^{2}\right] ,
\end{eqnarray*}%
we have%
\[
\frac{1}{2\phi }\nabla ^{b}\phi \left[ 2w\nabla ^{2}\phi -\frac{w}{\phi }%
\left( \nabla \phi \right) ^{2}\right] =\frac{\eta }{4\pi }\nabla
^{b}R+\nabla ^{b}h.
\]%
Using the equations of motion (\ref{em}), we obtain an integrability
condition%
\begin{equation}
\frac{\eta }{4\pi }\nabla ^{b}R+\nabla ^{b}h+\frac{1}{2}\nabla ^{b}\phi R=0.
\label{ic}
\end{equation}%
Obviously, this condition can be solved if%
\begin{equation}
\eta =2\pi \phi ,\;h=-\frac{\phi R}{2}.  \label{yh}
\end{equation}%
Substituting expressions (\ref{yh}) into equation (\ref{TR}), we finally
give the equation of state%
\begin{eqnarray*}
T^{mab}+T^{\phi ab} &=&\phi R^{ab}-\nabla ^{a}\nabla ^{b}\phi +g^{ab}\nabla
^{2}\phi -g^{ab}\frac{\phi R}{2} \\
&=&\phi G^{ab}-\left( \nabla ^{a}\nabla ^{b}\phi -g^{ab}\nabla ^{2}\phi
\right)
\end{eqnarray*}%
which is just the correct field equation (\ref{fe}). Also $\eta =2\pi \phi $
results that the entropy expression $S=\eta A$ is identical with the
expected black hole entropy $S=\frac{A}{4G_{eff}}$ in Brans-Dicke gravity
\cite{Cai2}.

The thermodynamic derivation for the equation of state can be easily
extended to general scalar-tensor theory of gravity with the Lagrangian%
\[
L=\frac{1}{2}F(\phi )R-\frac{1}{2}\left( \nabla \phi \right) ^{2}-V(\phi
)+L_{m},
\]%
In this case, gravitational field equation and $T_{ab}^{e}$ have similar
forms as (\ref{fe}) and (\ref{Te}), instead that $F(\phi )$ plays the role
of the gravitational constant. The equation of motion is%
\begin{equation}
\nabla ^{2}\phi -V^{\prime }(\phi )+\frac{1}{2}F^{\prime }(\phi )R=0,
\label{em2}
\end{equation}%
and the energy--momentum tensor of scalar field
\begin{equation}
T_{ab}^{\phi }=\nabla _{a}\phi \nabla _{b}\phi -g_{ab}\left( \frac{1}{2}%
\left( \nabla \phi \right) ^{2}+V(\phi )\right) .  \label{tf2}
\end{equation}%
Assuming conservation of matter, and using the covariant divergence of $%
T_{ab}^{\phi }$ (\ref{tf2})%
\[
\nabla _{a}T^{\phi ab}=\nabla ^{b}\phi \left( \nabla ^{2}\phi -V^{\prime
}\right)
\]%
and the equation of motion (\ref{em2}), the integrability condition (\ref{s2}%
) can be solved if $\eta =2\pi F,\;h=-\frac{FR}{2}$. Substituting these
functions into equation (\ref{TR}), one can obtain the correct field
equation.

In summary, we have studied the Rindler space-time thermodynamics in the
scalar-tensor gravity. We have shown that the gravitational field equation
can be derived as an equation of state of local equilibrium thermodynamics.
It is an alternative treatment to reinterpretate of the nonequilibrium
correction introduced in scalar-tensor gravity \cite{Cao,Akbar}. This work
has generalized the approach developed in \cite{Elizalde} for $f(R)$
gravity. It is known that $f(R)$ gravity can be treated as a special
scalar-tensor theory by introducing the scalar field $\phi =R$ and potential
$V=\phi f^{\prime }-f$ and choosing the Brans-Dick parameter $\omega =0$
(see \cite{Faraoni} for a review). We have shown the consistence between two
theories on the thermodynamic aspect, which is not affected by the concrete
potential and parameter. We have introduced a more general local entropy
expression as done in $f(R)$ gravity \cite{Elizalde}, where the
proportionality constant between the horizon area and entropy is replaced
with a field-depended function. Since this is motivated by the black hole
entropy expression in scalar-tensor gravity, we suspect that the entropy of
local Rindler horizon for the more generalized gravity should be modified as
corresponding black hole entropy. This is consistent with the entropy
expressions of cosmological apparent horizon in different gravity theories
\cite{Cao,Gong}. Our equilibrium thermodynamics is constructed by assuming
the conserved energy of matter, which suggests the energy exchange between
the matter and scalar field may lead to the nonequilibrium Rindler
spacetime. We have also assumed the energy--momentum tensor of scalar field
with nonminimal coupling to gravitation is just the one with minimal
coupling, which means that the effective energy $T_{ab}^{e}$ can \textit{not}
feel the heat flow across the horizon. This implies that we can not simply
deal with thermodynamicas of effective energy as the one of normal matter.
We expect that it may be helpful to understand the negative entropy of
phantom dark energy \cite{Nojiri} which may be an effective energy in
modified gravity theories.

\section*{Acknowledgements}

S. F. Wu is thankful to Sergei D. Odintsov for helpful discussions. This
work was supported by the NSFC under Grant Nos. 10575068 and 10604024, the
Shanghai Research Foundation No. 07dz22020, the CAS Knowledge Innovation
Project Nos. KJcx.syw.N2, the Shanghai Education Development Foundation, and
the Innovation Foundation of Shanghai University.

\end{document}